# Medical Aid for Automatic Detection of Malaria

Pramit Ghosh[1], Debotosh Bhattacharjee[2], Mita Nasipuri[2] and Dipak Kumar Basu [2],

[1] Department of Computer Science & Engineering, RCC Institute of Information Technology, Kolkata 700015, India
{ pramitghosh2002@yahoo.co.in }

[2] Department of Computer Science. & Engineering. Jadavpur University, Kolkata 700032, India
{ debotoshb@hotmail.com, mitanasipuri@gmail.com, dipakkbasu@gmail.com }

**Abstract.** The analysis and counting of blood cells in a microscope image can provide useful information concerning to the health of a person. In particular, morphological analysis of red blood cell's deformations can effectively detect important disease like malaria. Blood images, obtained by the microscope, which is coupled with a digital camera, are analyzed by the computer for diagnosis or can be transmitted easily to clinical centers than liquid blood samples. Automatic analysis system for the presence of Plasmodium in microscopic image of blood can greatly help pathologists and doctors that typically inspect blood films manually. Unfortunately, the analysis made by human experts is not rapid and not yet standardized due to the operators' capabilities and tiredness. The paper shows how effectively and accurately it is possible to identify the Plasmodium in the blood film. In particular, the paper presents how to enhance the microscopic image and filter out the unnecessary segments followed by the threshold based segmentation and recognize the presence of Plasmodium. The proposed system can be deployed in the remote area as a supporting aid for telemedicine technology and only basic training is sufficient to operate it. This system achieved more than 98% accuracy for the samples collected to test this system.

## 1  Introduction

Malaria is a mosquito-borne infectious disease of humans caused by a parasite called Plasmodium. It is widespread in tropical and subtropical regions, including much of Sub-Saharan Africa, Asia and America. In the human body, the parasites multiply in the liver (exoerythrocytic phase) [1], and then infect red blood cells (erythrocytic phase). Symptoms of malaria include fever, headache, and vomiting, and usually appear between 10 and 15 days after the mosquito bite. If not treated, malaria can quickly become life-threatening by disrupting the blood supply to vital organs [2]. The accepted laboratory practice for the diagnosis of malaria is the preparation and microscopic examination of blood films stains generated by Giemsa's solution. Figure

1 shows 1125X Magnified blood sample, which reveals the presence of two Plasmodium vivax parasites; an immature form on the left, and another in a mature form on the right.

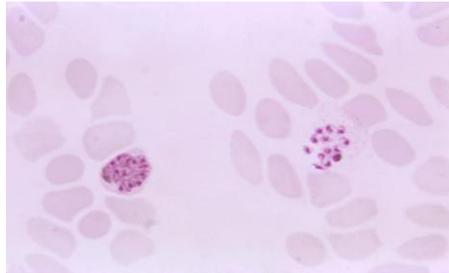

**Fig. 1.** 1125X Magnified blood sample, with two Plasmodium vivax parasites.

Four species of Plasmodium parasites can infect human beings and also they can utilize human beings as temporary repository in transmitting the disease malaria from one sick individual to another healthy person. Severe disease is largely caused by Plasmodium falciparum [3] and it is also responsible for about 90% of the deaths from malaria. Malaria caused by Plasmodium vivax, Plasmodium ovale and Plasmodium malariae is generally a milder disease that is rarely fatal [4].

The World Health Organization (WHO) reports that over 780,000 people died of malaria in 2009 [5], most of them are children, under the age of five. At a particular time, an estimated 300 million people are said to be infected with at least one of these Plasmodium species. Sometimes it may so happens that malaria becomes an epidemic in the particular time of the year and that generally occurs during rainy season. So, there is a great need for the development of effective treatments for decreasing the yearly mortality and morbidity rates. Once an epidemic like situation emerges, it becomes very difficult to arrange sufficient pathologist and equipments for diagnosis of the Malaria especially in economically backward areas. Some test kits exist like "Malaria home test kit" [6]. But cost of such kit is much higher. As a matter of fact, it is difficult to place such kits in every health care unit. The objective of this work is to design a low cost device which is capable to detect Plasmodium species from blood film images to speed up the diagnosis process.

The rest of the paper is arranged as follows: section 2 describes the proposed system; section 3 presents the results and performance of the system and section 4 concludes the work along with discussions on the future scope of this work.

## 2   System Detail

The system is explained with the help of a block diagram, shown in figure 2, and all the steps of the system are described next.

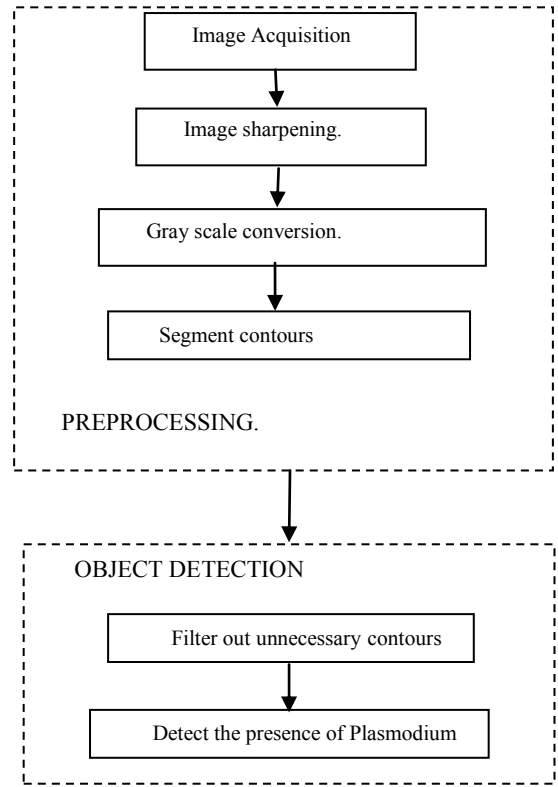

**Fig. 2.** Block diagram of the system.

### 2.1 Image Acquisition

Image Acquisition is the first step of the system. Digitized images of the blood samples on the slides are acquired with a CCD [7] camera which is mounted upon the microscope. For getting multiple images of a single sample, the glass slide movement is required and it is controlled by two stepper motors [8] in the horizontal and vertical direction shown in Figure 3.

### 2.2 Image Enhancement

The images obtained from the CCD camera are not of good quality. Laplacian Filter [9] is used to sharpen the edges of the objects in the image.
The Laplacian value for a pixel is denoted by $\nabla^2 f(x, y)$ and it is defined as..

$$\nabla^2 f(x, y) = \frac{\partial^2}{\partial x^2} f(x, y) + \frac{\partial^2}{\partial y^2} f(x, y) \qquad (1)$$

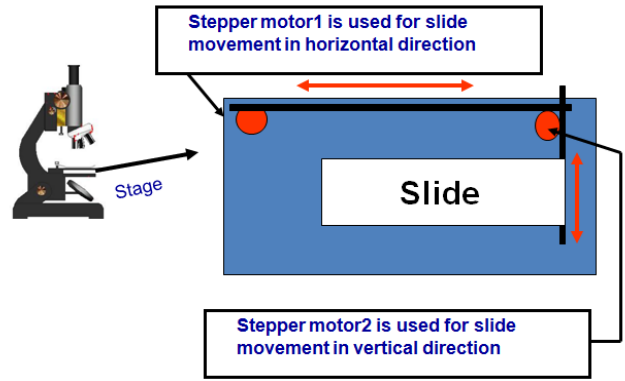

**Fig. 3.** Slide control through stepper motor.

This expression is implemented at all points (x,y) of the image through convolution. The Laplacian filter is applied separately on Red, Green and Blue components of the colour images obtained from the CCD camera. After that, the images are converted into gray scale image by simple average of three components namely Red, Green, and Blue.

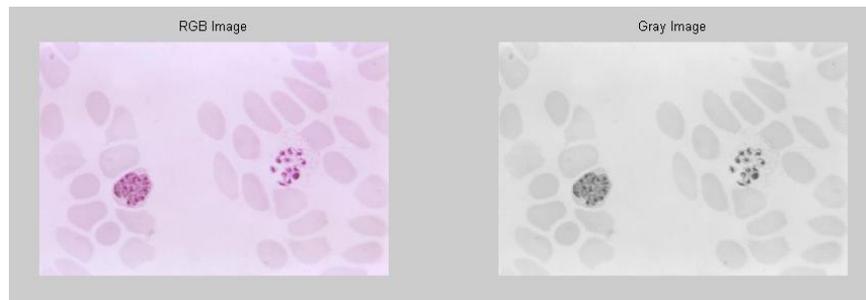

**Fig. 4** RGB Image and its corresponding gray scale image.

### 2.3  Segmentation

By analyzing gray scale images it is inferred that the area of the image occupied by the Plasmodium has low gray value and high intensity variation with respect to other area of the image. For the benefit of the implementation, all the gray images are converted into corresponding negative images, in which light areas appear dark and vice versa. As a result, in the inverted images the area occupied by the Plasmodium will have high gray value with respect to other area of the image.

All the input blood film images are not illuminated in the same scale. So inverted gray scale image illumination is not in the same scale; this might cause problem in threshold based region segmentation. Uniform scale conversion is required to overcome this problem. It is implemented by subtracting a calculated value from each pixel points of the inverted image. The value which is proposed to be subtracted is the 45% of the average gray value of the pixels of the inverted image.

Trial and error method reveals that the subtraction is only applicable if the difference between the maximum gray scale value and average gray scale value of the inverted gray image is greater than a threshold value, which is obtained by statistical data analysis. Image histogram is used to find out the maximum value; otherwise noise with high gray value might claim the maximum value.

Algorithm-1: explains the scaling process.
Step 1: Calculate the mean value of the gray value of the inverted image.
Step 2: Calculate the histogram and find out the average of a group of pixels with top 1/80 of total number of pixels in the inverted image. This process will eliminate the chance of noise pixel with high gray value getting selected.
Step 3: Find out the difference between two values supplied by step 1 and 2.
Step 4: If value obtained in step3 is greater than predetermined threshold value then subtract 45% of the average value obtained in step 1. This is applied to each pixel points of the inverted image.
Step 5: Stop.

After applying algorithm-1, the inverted gray image is converted into binary image using a threshold value. This threshold value varies from image to image. The threshold value calculation [12] algorithm is given next.

Algorithm-2: Calculation of threshold value.

Step 1: Select an initial estimate for T (T=threshold value). The initial value for T is the average gray level of the image
Step 2: Segment the images using T. This will produce two groups of pixels: consisting of all the pixels with gray level values > T called as G1 and consisting of pixels with values <= T called as G2.
Step 3: Compute the average gray level values $\mu1$ and $\mu2$ for the pixels in regions G1 and G2.
Step 4: Compute a new threshold value: $T = 0.5 * (\mu1+ \mu2)$.
Step 5: Repeat steps 2 through 4 until the difference in T in successive iterations is smaller than a predefined parameter $T_0$.
Step 6: Stop.

**2.4** Filter out Unnecessary Contours

The binary image has unnecessary contours that are basically noise. This small noise contours are eliminated by closing, which is dilation[13] followed by erosion [13]. Closing is able to remove unnecessary contours which are small in size but fails to eliminate unnecessary contours of big in size. Figure 5 shows the binary image and its corresponding binary image after removal of small unnecessary contours.

The unnecessary contours which have considerable size are indicating the regions of red blood cells. The difference between red blood cells and Plasmodium is that red blood cells have smooth surface whereas Plasmodium has rough surface area. The gradient operator is helpful to distinguish the picture segment of red blood cell and Plasmodium. The gradient at the center point in a neighborhood is computed as given in [12]

The gradient operator is applied on the inverted gray image and then the output is converted into binary using a threshold value. Figure 6A shows the output along with merging the two binary images given in Figure 6B. The cluster of white region denotes the surface is not smooth.

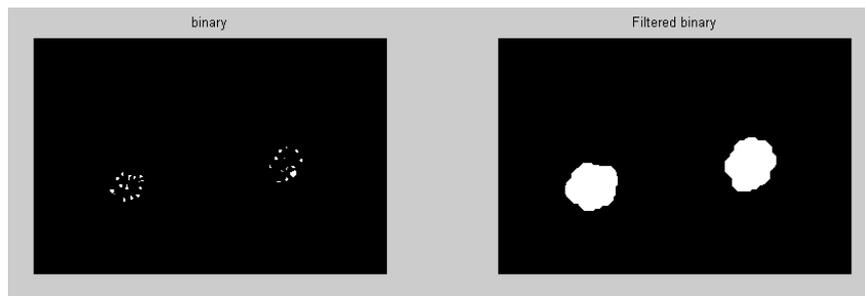

**Fig. 5A** Filtered Binary image             **Fig, 5B** After removal of small contours from 5A.

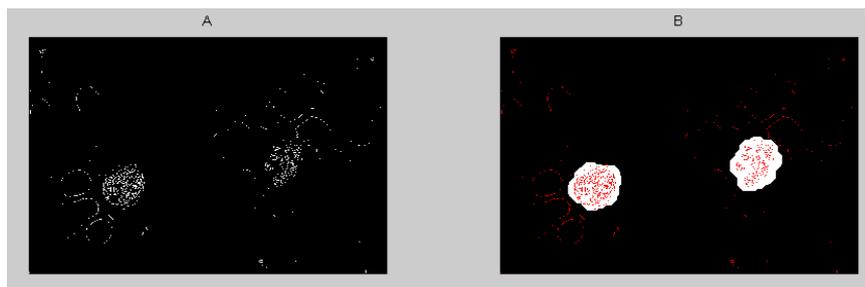

**Fig. 6A** Binary image after gradient operator             **Fig. 6B** Overlapping of figure 5B and 6A.

### 2.5 Detection of the Presence of Plasmodium

By comparing two sets of binary images, obtained from previous section, the presence of Plasmodium is determined. This is done by the following algorithm.
Algorithm-3: To find out a valid contour of Plasmodium.

Step1: Apply label matrix technique in the binary image obtained from the inverted gray image, and store the output matrix in LB variable. LB will have the same dimension as the binary image.
Step 2: count = maximum integer value stored in LB; So "count" will contain the number of contours in the input image.
Step 3: index = 1
Step 4: val = (total number of points, where value is 1, in the binary image, which is obtained after applying gradient operator) / (total number of points in the binary image)
Step 5 : Find the coordinates of the pixels of the LB where value of the pixel is equal to index;
Step 6: local_Value = (find out the number of points, where value is 1, in the binary image, which is obtained after applying gradient operator and whose coordinates are selected in step 5.) / (total number of coordinates are selected in step 5.)
Step 7: if local_Value > >val
  then contour of a Plasmodium is found
  Else Not a valid plasmodium contour
Step 8 : index = index + 1;
Step 9: Repeat step 5 to step 8 until index > count
Step 10 : Stop

## 3 Simulation and Results

For simulation MatLab 7.1 [14] is used. The image shown in Figure 1 is fed as an input. After analysis, system finds the presence of Plasmodium. Output is shown next.
```
Value = 0.0088
local_Value = 0.1559
ans = Plasmodium found
local_Value = 0.1023
ans = Plasmodium found
```

For testing purpose 160 samples were collected and out of which 157 samples have shown accurate results. So the accuracy of the system is found to be 98.125%.

## 3 Conclusion

In this paper one novel approach for detection a malarial parasite, called Plasmodium, is proposed. This system is cheaper than other Malaria test kit. This system does not require any special technical skill. So, this can be used by the people of remote places with very basic level of education. It may reduce the probability of wrong treatment which happens due to non-availability of diagnosis systems in remote and economically backward areas. As a next phase authors are trying to design an integrated system for diagnosis of diseases due to such parasites.

**Acknowledgment**

Authors are thankful to Dr. Abhijit Sen and Dr Soumendu Datta for providing pathological data and the "DST-GOI funded PURSE programme", at Computer Science & Engineering Department, Jadavpur University, for providing infrastructural facilities during progress of the work.